\documentclass[twocolumn]{aastex631}

\received{2023 June 11}
\accepted{2023 September 12}
\submitjournal{PASP}

\shorttitle{}
\shortauthors{Kov\'acs-Stermeczky \& Vink\'o}

\begin{document}

\title{Fitting optical light curves of Tidal Disruption Events with TiDE}

\correspondingauthor{Zs. Kov\'acs-Stermeczky}
\email{stermeczky.zsofia@csfk.org}

\author[0000-0002-5655-0154]{Zs\'ofia V. Kov\'acs-Stermeczky}
\affiliation{ELTE E\"otv\"os Lor\'and University, Institute of Physics and Astronomy, P\'azm\'any P\'eter s\'et\'any 1/A, Budapest, 1117 Hungary}

\affiliation{ Konkoly Observatory,  CSFK, MTA Centre of Excellence, Konkoly Thege M. \'ut 15-17, 
Budapest, 1121, Hungary}

\author[0000-0001-8764-7832]{J\'ozsef Vink\'o}
\affiliation{ Konkoly Observatory,  CSFK, MTA Centre of Excellence, Konkoly Thege M. \'ut 15-17, 
Budapest, 1121, Hungary}
\affiliation{ELTE E\"otv\"os Lor\'and University, Institute of Physics and Astronomy, P\'azm\'any P\'eter s\'et\'any 1/A, Budapest, 1117 Hungary}
\affiliation{Department of Experimental Physics, University of Szeged, D\'om t\'er 9, Szeged, 6720, Hungary}
\affiliation{Department of Astronomy, University of Texas at Ausin, 2515 Speedway Stop C1400, Austin, TX, 78712-1205, USA}

\begin{abstract}
A Tidal Disruption Event (TDE) occurs when a supermassive black hole tidally disrupt a nearby passing star. The fallback accretion rate of the disrupted star may exceed the Eddington limit, which induces a supersonic outflow and a burst of luminosity, similar to an explosive event. Thus, TDEs can be detected as very luminous transients, and the number of observations for such events is increasing rapidly.
In this paper we fit $20$ TDE light curves with {\tt TiDE}, a new public,  object-oriented code designed to model optical TDE light curves. We compare our results with those obtained by the popular {\tt MOSFiT} and the recently developed {\tt TDEmass} codes, and discuss the possible sources of differences.
\end{abstract}


\section{Introduction}

When a star approaches a supermassive black hole (SMBH) more than a critical distance (called as the tidal radius), the tidal forces from the SMBH could partially or totally disrupt the nearby passing star. This is called Tidal Disruption Event (TDE). In the first approximation, about half of the stellar debris will leave the system while the other half will be bound to the SMBH \citep{1988Natur.333..523R}. Such an event was first predicted theoretically by Hills \citep{1975Natur.254..295H}. More recent results are summarized in e.g. \citet{2020SSRv..216..124V} or \citet{2021ARA&A..59...21G}.

Nowadays TDEs have become a popular part of recent astrophysical literature, thanks to the increasing number of observations produced mostly by large sky surveys like the All-Sky Automated Survey for Supernovae (ASASSN), Sloan Digital Sky Survey (SDSS) or the Zwicky Transient Facility (ZTF) \citep[e.g.][]{2008ApJ...678L..13K, 2011ApJ...741...73V, 2011ApJ...740...85W, 2012ApJ...749..115W, 2014MNRAS.445.3263H, 2016MNRAS.455.2918H, 2019ApJ...872..198V, 2021ApJ...908....4V}. As a result, a growing number of photometric as well as spectroscopic observations are available for $\sim 200$ TDEs at present. 

The physical processes behind the TDE phenomenon are quite complex, and it is still not clear exactly which processes are responsible for the observed luminosity. Several implementations of TDE light curves contain at least two components: an accretion disk around the SMBH and a super-Eddington wind/outflow originating near the bottom of the accretion disk  \citep[e.g.][]{Lodato_Rossi11}. There are many other possible sources, like the reprocessing of the high-energy radiation from the disk to optical photons by the wind material \citep[e.g.][]{Guillochon14} or by the unbound material \citep[e.g.][]{Strubbe09}, the effect of shocks due to infalling stream-disk interactions \citep{2020MNRAS.495.1374B, 2022arXiv220610641S}, or shocks near the apocenter of the stream \citep{2015ApJ...804...85S, 2015ApJ...806..164P}.

These sources can be modeled by different assumptions with varying level of complexity; for example, the disk could be a simple thin disk with a time-independent, constant outer radius ($r_{\rm out}$) \citep{Lodato_Rossi11}, while other models adopt a spreading disk scenario \citep{2016MNRAS.455..859S}. 
Many papers explored these models via semi-analytic calculatons, while others preferred more sophisticated hydrodynamical simulations \citep{L09, 2011MNRAS.413.1623D, 2013ApJ...767...25G, 2016MNRAS.461..948M, Golightly_2, Golightly_1, nixon21}.

How the star can approach the SMBH close enough is still an unanswered question, but there are several theoretical arguments e.g. from SMBH binaries \citep{2005MNRAS.358.1361I, 2009ApJ...697L.149C,2011ApJ...729...13C, 2015MNRAS.451.1341L} to standard two-body relaxation \citep{1976MNRAS.176..633F, 1999MNRAS.309..447M, 2016MNRAS.455..859S}. From these scenarios the orbit of the nearby passing star could also be eccentric instead of parabolic. If the star survives such an approach, it can produce periodic outbursts during subsequent passages near the pericenter \citep{2013ApJ...777..133M, 2022ApJ...927L..25N}.

Several analytical or semi-analytical models exist that predict the observables (mostly the light curve) of TDEs, which can be used to constrain some physical parameters, like the mass of the SMBH and the destroyed star. 
One of them, {\tt TDEmass} \citep{TDEmass}, estimates these parameters only from the bolometric luminosity at peak. On the contrary, two other codes, {\tt MOSFiT} \citep[][M19 hereafter]{Mosfit} and {\tt TiDE} \citep[][KV23 hereafter]{TiDE}, model the whole observed light curve in optical bands. 

Both {\tt TiDE} and {\tt MOSFiT} predict the parameters of TDEs based on the early-time light curve (before and shortly after the peak). However, there are other approaches that are based on late-time observations \citep[e.g.][]{2019ApJ...878...82V, 2023ApJ...948...68Z}. \citet{2019ApJ...878...82V} studied the late-time UV observations and found that they cannot be explained by the same model as the early-time light curve (i.e. reprocessing or circularization shocks), but they can be modeled quite well by a viscously spreading, unobscured accretion disk.

Along with the increasing number of TDE observations, a possibility emerged for studying TDEs based on their spectroscopic properties. For example,  \citet{2022MNRAS.515.5604N} defined three different (empirical) classes: $i)$ only hydrogen lines (TDE-H), $ii)$ only helium lines(TDE-He), and $iii)$ the mixture of H, HeII and often NIII/OIII features (TDE-H+He). They tested the relationship between these classes and their parameters based on $32$ TDEs, whose light curves were fitted by {\tt MOSFiT}. They found that TDE-He events are probably total disruptions, while TDE-H events may be less complete disruptions than TDE-H+He. They also found an approximatly linear correlation between the mass of the SMBH and the radiative efficiency of the accretion onto it.

Attempts to finding TDEs of white dwarf stars around intermediate-mass black holes also appeared in recent literature.  \citet{2023arXiv230214070G} systematically queried the entire ZTF alert stream to find, as they called, Ia-TDE candidates. They restricted their search to events in dwarf galaxies, and found six candidates. They compared the observations to a theoretical light curve and a spectroscopic model created by \citet{2016ApJ...819....3M} to identify the best candidates.

Still, the number of TDEs with well-modeled observations and tightly constrained physical parameters is quite low. 
In this paper we apply the {\tt TiDE} \citep[][, KV23 hereafter]{TiDE} code for fitting $20$ well-sampled TDE light curves, then we compare the results with previously published {\tt MOSFiT} and {\tt TDEmass} model fits. Beside using the recently developed {\tt TiDE} code for TDEs with different observational characteristics, our motivation is to test the model-dependency of the most common physical parameters of TDEs, i.e. the mass of the SMBH and the disrupted star. 
This paper is organized as follows.
In Section \ref{sec:tde_models} we briefly summarize the physics of the adopted TDE model. The observational sample and the selection criteria are described in Section \ref{sec:select_obj}, while the results and the discussion can be found in \ref{sec:results}.

\section{TDE models}
\label{sec:tde_models}

In this section we briefly describe the semi-analytic TDE models that we consider in the present paper. 

\subsection{{\tt TiDE}}
TiDE calculates the quasi-monochromatic luminosity of a TDE based on the formulae given by \citet{Strubbe09}, \citet{L09} (L09 hereafter) and \citet{Lodato_Rossi11}. The light curve contains primarily emission from two components: an accretion disk and a super-Eddington wind. This simple picture can be extended to a more complex one by adding the following process: the bolometric luminosity of the disk component could be reprocessed by the wind material, which may increase both the temperature and the luminosity of the wind component. {\tt TiDE} contains a simple model of this reprocessing effect, which assumes that the wind reprocesses the bolometric disk luminosity with a constant $\epsilon_{\rm rep}$ efficiency, and it increases the photospheric temperature accordingly.

{\tt TiDE} contains many different models for the key parameters (see KV23 for more details): the fallback accretion rate ($\dot{M}_{fb}$) and the fallback timescale ($t_{\rm min}$).
Here we use the fallback accretion rate as defined by L09 with a main sequence star having either  $n=3$ or $n=3/2$ polytropic indices. The $t_{\rm min}$ timescale
also has several different built-in prescriptions. 
We use a corrected $t_{\rm min}$ that calibrates the tidal radius for full disruptions to those from hydrodinamical simulations. For this calibration we introduce a correction factor, $\beta_d$ (see KV23 for details). 
In this paper we use this corrected $t_{\rm min}$ with $\beta_d = 1.85$ \citep{Guillochon14} for $n = 3$ and $\beta_d = 1.2$ (KV23) for $n = 3/2$, corresponding to full disruption of the star. 

There are some other, more technical parameters, like the mass ratio of the super-Eddington outflow and the accreting mass ($f_{\rm out}$), or the diffusion timescale for photons propagating through the optically thick wind. In the present calculations we use a time-dependent $f_{\rm out}$ parameter, and photon diffusion is also taken into account (see KV23 for more details).  

In order to fit multiband light curves, we feed the output of {\tt TiDE} into the {\tt MINIM} fitting code \citep{Minim}, which uses a Monte-Carlo $\chi^2$-minimalization method to find the best-fit model to each observed light curve.

The penetraction factor $\beta = r_t/r_p$,  where $r_t$ is the tidal radius, while $r_p$ is the pericenter distance of the stellar orbit,
in the current implementation of {\tt TiDE} affects only the outer radius of the disk, which causes only a minor perturbation in the light curve near the peak. Here we does not attempt to fit this parameter. Instead, in all cases we assume $\beta = 1$.
In addition, $\beta$ must have an effect on the calculated fallback accretion rate as well, especially in partial disruptions that are not implemented in the current version of TiDE. 
Another reason for using $\beta = 1$ is that within the framework of {\tt TiDE} changing $\beta$ would alter only the disk luminosity, which would affect only the $\epsilon_{\rm rep}$ parameter.
Since $\beta$ and $\epsilon_{\rm rep}$ are strongly correlated parameters, the simultaneous fitting of them could result in unphysical values for both parameters. Therefore, all the best-fit models presented in this paper are based on the constraint of $\beta = 1$.  

\subsection{{\tt MOSFiT}}

The TDE module in {\tt MOSFiT} (M19) is a popular model for representing single-band light curves of TDE events. {\tt MOSFiT} uses nine free parameters including the mass of the black hole and the star. The fallback accretion rate is derived from hydrodynamically simulated fallback accretion rates with different impact parameters ($\beta$) for $M_* = 1$ and $M_6 = 1$ \citep{2013ApJ...767...25G}, and analytic relations are used to scale the fallback accretion rates into other $M_*$ and $M_6$ masses. These simulations were also computed for low $\beta$ parameters, therefore, in principle, {\tt MOSFiT} can be applied for both partial and total disruptions. The code creates a bolometric luminosity assuming that a fraction of the energy from the fallback is converted into radiation, 
then transforming it into multicolor information from the assumption that the radiation is reprocessed by an extended photosphere.  An important restriction built-in {\tt MOSFiT} is that the fallback accretion rate cannot exceed the Eddington limit for a given SMBH mass ($M_6$). Thus, the {\tt MOSFiT} model does not contain the super-Eddington wind component, which may have a non-negligible effect on the best-fit parameter values.

\subsection{{\tt TDEmass}}

{\tt TDEmass} \citep{TDEmass} estimates the black hole and stellar masses from the observed bolometric luminosity and temperature at the peak. This method uses a different physical model than the previous two codes: it assumes that the circularization of the debris material is slow, and the source of the the observed optical/UV radiation is the shock in the intersecting/colliding debris streams.

\section{Selected objects}
\label{sec:select_obj}

\begin{table*}
    \begin{rotatetable*}
    \centering
    \begin{center}
    \caption{Selected objects and their sources. The source numbers corresponds to the first and second {\tt MOSFiT} and  {\tt TDEmass} references, respectively.}
    \label{tab:obj_masses}
    \begin{tabular}{cccccccc}
    \hline
    \hline
    Object name & source & {\tt MOSFiT} $M_6$ & {\tt TDEmass} $M_6$ & {\tt TiDE} $M_6$ & {\tt MOSFiT} $M_*$ & {\tt TDEmass} $M_*$ & {\tt TiDE} $M_*$\\
    \hline
    PS1-10jh & 1, 4, 5 & $17^{+2}_{-1}$, $10^{+0.96}_{-1.29}$ & $2.3^{+0.4}_{-0.5}$ & $ 7.66 \pm 0.66 $ & $0.101^{+0.002}_{-0.002}$, $0.41^{+0.09}_{-0.16}$ & $1.8^{+1.1}_{-0.5}$ & $ 27.12 \pm 1.73 $\\
    PS1-11af & 1, 4, 5 & $3.7^{+0.5}_{-0.4}$, $2.82^{+0.34}_{-0.25}$ & $8.9^{+1.8}_{-2.7}$ & $ 1.46 \pm 0.08 $ & $0.101^{+0.009}_{-0.003}$, $0.96^{+0.03}_{-0.06}$ & $1.8^{+0.6}_{-0.4}$ & $ 17.09 \pm 0.29 $\\    
    PTF09ge & 1, 4, 5 & $3.6^{+0.8}_{-0.5}$, $2.95^{+0.28}_{-0.26}$ & $4.1^{+0.9}_{-1.1}$ & $ 1.20 \pm 0.03 $ & $0.10^{+0.07}_{-0.01}$, $0.08^{+0.01}_{-0.00}$ & $1.8^{+1.8}_{-0.6}$ & $ 15.09 \pm 0.04 $\\
    iPTF16fnl & 1, 4 & $1.7^{+0.2}_{-0.2}$, $0.79^{+0.33}_{-0.10}$ & - & $ 0.21 \pm 0.00 $ & $0.101^{+0.008}_{-0.004}$, $0.98^{+0.03}_{-0.87}$ & - & $ 0.69 \pm 0.01 $\\
    ASASSN-14li & 1, 4 & $9^{+2}_{-3}$, $10^{+2.02}_{-2.24}$ & - & $ 0.51 \pm 0.11 $ & $0.2^{+0.1}_{-0.1}$, $0.18^{+0.07}_{-0.05}$ & - & $ 4.41 \pm 0.44$\\
    ASASSN-15oi & 1, 4 & $4^{+1}_{-1}$, $5.37^{+0.25}_{-0.24}$ & - & $ 17.08 \pm 0.64 $ & $0.11^{+0.04}_{-0.01}$, $0.12^{+0.11}_{-0.02}$ & - & $ 22.00 \pm 0.18 $\\
    ASASSN-14ae & 1, 4 & $1.3^{+0.1}_{-0.1}$, $1.35^{+0.16}_{-0.12}$ & - & $ 0.76 \pm 0.13 $ & $1.00^{+0.02}_{-0.02}$, $0.99^{+0.14}_{-0.08}$ & - & $ 5.23 \pm 0.86 $\\
    iPTF16axa & 1, 4 & $2.5^{+1.3}_{-0.9}$, $19.50^{+5.62}_{-5.04}$ & - & $ 2.82 \pm 0.23 $ & $1.0^{+0.8}_{-0.2}$, $0.33^{+0.17}_{-0.07}$ & - & $ 11.15 \pm 1.19 $\\
    D1-9 & 1, 4 & $66^{+7}_{-10}$, $6.17^{+4.80}_{-3.28}$ & - & $ 16.44 \pm 3.94 $ & $7^{+5}_{-3}$, $0.28^{+0.21}_{-0.14}$ & - & $ 10.04 \pm 1.16 $\\
    D3-13 & 1, 4 & $30^{+3}_{-3}$, $10^{+6.22}_{-3.97}$ & - & $ 23.68 \pm 0.86 $ & $7^{+17}_{-4}$, $0.36^{+0.79}_{-0.12}$ & - & $ 60.00 \pm 0.68 $\\
    AT2018hco & 2, 4, 2, 5 & $3.16^{+4.08}_{-0.92}$ , $4.37^{+1.66}_{-1.27}$ & $3.98^{+0.19}_{-0.18}$, $3.3^{+0.1}_{-0.2}$ & $ 7.57 \pm 0.83 $ &  $1.54^{+0.84}_{-0.63}$, $0.08^{+0.01}_{-0.01}$ & $2.4^{+0.75}_{-0.39}$, $2.6^{+0.7}_{-0.6}$ & $ 25.88 \pm 1.81 $\\
    AT2018lna & 2, 2, 5 & $6.76^{+3.71}_{-1.75}$ & $1.29^{+0.03}_{-0.03}$, $1.3^{+0.3}_{-0.3}$ & $1.38 \pm 0.35$  & $2.85^{+2.82}_{-1.76}$ & $5.50^{+3.00}_{-1.40}$, $5.7^{+3.5}_{-1.8}$ & $ 7.41 \pm 0.50 $\\
    AT2018hyz & 2, 4, 2 & $3.80^{+0.46}_{-0.49}$, $3.72^{+0.36}_{-0.33}$ & $5.89^{+0.28}_{-0.27}$ & $ 0.05 \pm 0.00 $ & $0.99^{+0.05}_{-0.06}$, $0.96^{+0.04}_{-0.05}$ & $4.60^{+13.00}_{-1.80}$ & $ 35.81 \pm 2.48 $\\
    AT2019azh & 2, 4, 2, 5 & $26.92^{+7.76}_{-13.73}$, $5.01^{+0.74}_{-0.75}$ & $2.19^{+0.05}_{-0.00}$, $0.74^{+0.09}_{-0.09}$ & $ 1.03 \pm 0.06 $ & $3.59^{+2.55}_{-0.95}$, $0.47^{+0.20}_{-0.07}$ & $3.60^{+0.70}_{-0.39}$, $13^{+2}_{-2}$ & $ 9.94 \pm 0.58 $\\
    AT2019ehz & 2, 4, 2, 5 & $6.03^{+1.22}_{-0.90}$, $2.19^{+0.21}_{-0.19}$ & $3.24^{+0.23}_{-0.22}$, $3.9^{+0.3}_{-0.3}$ & $ 1.56 \pm 0.07 $ & $9.81^{+2.89}_{-3.41}$, $0.10^{+0.00}_{-0.00}$ & $1.20^{+0.14}_{-0.08}$, $1.4^{+0.1}_{-0.1}$ & $ 11.65 \pm 0.25 $\\
    AT2019meg & 2, 4, 2, 5 & $4.79^{+9.34}_{-1.32}$, $3.31^{+0.49}_{-0.43}$ & $3.47^{+0.00}_{-0.08}$, $2.6^{+0.1}_{-0.1}$ & $ 0.71 \pm 0.04 $ & $0.96^{+1.48}_{-0.54}$, $0.1^{+0.00}_{-0.00}$ & $3.10^{+0.34}_{-0.20}$, $4.3^{+1.1}_{-0.7}$ & $ 11.74 \pm 0.22 $\\
    AT2020pj & 2, 2 & $95.50^{+4.50}_{-6.37}$ & $2.88^{+0.35}_{-0.31}$ & $ 0.15 \pm 0.02 $ & $10.37^{+3.78}_{-4.77}$ & $0.59^{+0.05}_{-0.05}$ & $ 6.24 \pm 0.53 $\\
    AT2020mot & 2, 2 & $4.68^{+2.24}_{-1.66}$ & $3.24^{+0.48}_{-0.28}$ & $ 5.75 \pm 0.56 $ & $1.01^{+1.50}_{-0.12}$ & $1.10^{+0.17}_{-0.10}$ & $ 24.52 \pm 2.63 $\\
    AT2020opy & 2, 2 & $7.08^{+3.15}_{-2.07}$ & $2.88^{+0.00}_{-0.00}$ & $ 2.28 \pm 0.18 $ & $2.69^{+1.64}_{-0.82}$ & $0.50^{+0.09}_{-0.02}$ & $ 20.25 \pm 1.19 $\\
    AT2020wey & 2, 3, 2 & $22.91^{+2.21}_{-1.53}$, $2.88^{+0.66}_{-0.54}$ & $0.43^{+0.00}_{-0.00}$ & $ 0.16 \pm 0.01 $ & $4.34^{+1.96}_{-1.53}$,$0.11^{+0.05}_{-0.02}$ & $0.48^{+0.02}_{-0.01}$ & $ 0.52 \pm 0.02 $\\
    \hline
    \end{tabular}
    \end{center}
    \tablecomments{References: (1) \citet{Mosfit}; (2) \citet{ZTF_30_TDE}; (3) \citet{2022arXiv220912913C}; (4) \citet{2022MNRAS.515.5604N}; (5) \citet{TDEmass}}
\end{rotatetable*}
\end{table*}

The selection procedure for the observational sample is based on the following criteria. First, we intend to compare our fitting parameters (especially the masses) to those given by {\tt MOSFiT} and {\tt TDEmass}. Thus, we primarily selected objects from the papers by M19, \citet{ZTF_30_TDE} and \citet{2022MNRAS.515.5604N}. 

The authors of M19 fitted $14$ TDE light curves with MOSFiT and the data were downloaded from github\footnote{https://github.com/astrocatalogs/tde-1980-2025}. 

The \citet{ZTF_30_TDE} paper is mostly based on ZTF optical ($g$, and $r$) photometry, but they were also supplemented by additional UV measurements. \citet{ZTF_30_TDE} fitted 30 TDEs with both {\tt MOSFiT} and {\tt TDEmass}, but present only the black hole and stellar masses in their Table 7. The photometric data used by them are available in the electronic version of the paper. 

The \citet{2022MNRAS.515.5604N} paper contains all objects from M19 and many other data from \citet{ZTF_30_TDE}, too. They used {\tt MOSFiT} for fitting their sample, and the best-fit parameters can be found in their 
Table~1. 

Our second selection criterion is to find objects that have optical photometric observations, the more the better.  It is also preferable to have both the rising and the declining part of the light curve covered by the observations. The final list of our observational sample that satisfies these two selection criteria can be found in Table \ref{tab:obj_masses}.

All downloaded photometric data were corrected for Milky Way extinction 
and for luminosity distance. In the case of observations from \citet{ZTF_30_TDE} the distances were calculated from redshift, assuming Planck18 cosmology \citep{Planck18}. 
For the other objects we adopted the same distance as in the original publication (M19).

During the fitting we used mainly the data from optical bandpass filters, because the current version of TiDE is applicable only for such bandpasses. All observations were horizontally shifted by an arbitrary value ($T_0$) close to the suspected moment of the disruption in order to replace the Julian dates with lower numbers. 
This is only a technical parameter, and it was automatically selected equal to the MJD of the first data point for each object. 
The fitting region was restricted to within a few hundred days before and after the disruption. There are multiple reasons for this restriction. First, the reprocessing of high-energy photons by the unbound stellar material could be significant at late phases, several hundred days after peak \citep{Strubbe09}. Although TiDE contains a simple model for the reprocessing by the wind material, it does not take into account this effect with the unbound particles at high radii. Second, the time-dependent $f_{out}$ parameter, as implemented in TiDE, is applicable only at phases near the peak, when the accretion can be super-Eddington. After that its definition is no longer valid. In TiDE, when $f_{\rm out}$ reaches 0.1, the code switches to use this value as a constant $f_{\rm out}$ for further calculations. It may introduce an
artificial break in the light curve, which is unphysical. In order to avoid the problems it might cause in the modeling, very late-phase data were omitted from the fitting.

\section{Results and their comparison with MOSFiT and TDEmass}
\label{sec:results}

In Table~\ref{tab:obj_masses} we present the best-fit SMBH and stellar masses computed with the three different methods, i.e. {\tt MOSFiT}, {\tt TDEmass} and {\tt TiDE}. Those events that were observed both before and after the luminosity peak have parameter estimates from all three methods, while for those that were observed only during the decline phase, the {\tt TDEmass} results are missing. Note that most of the objects were analyzed in more than one paper previously, and we present all of those earlier results in Table \ref{tab:obj_masses}.

\begin{figure*}
    \centering
    \includegraphics[width=0.8\textwidth]{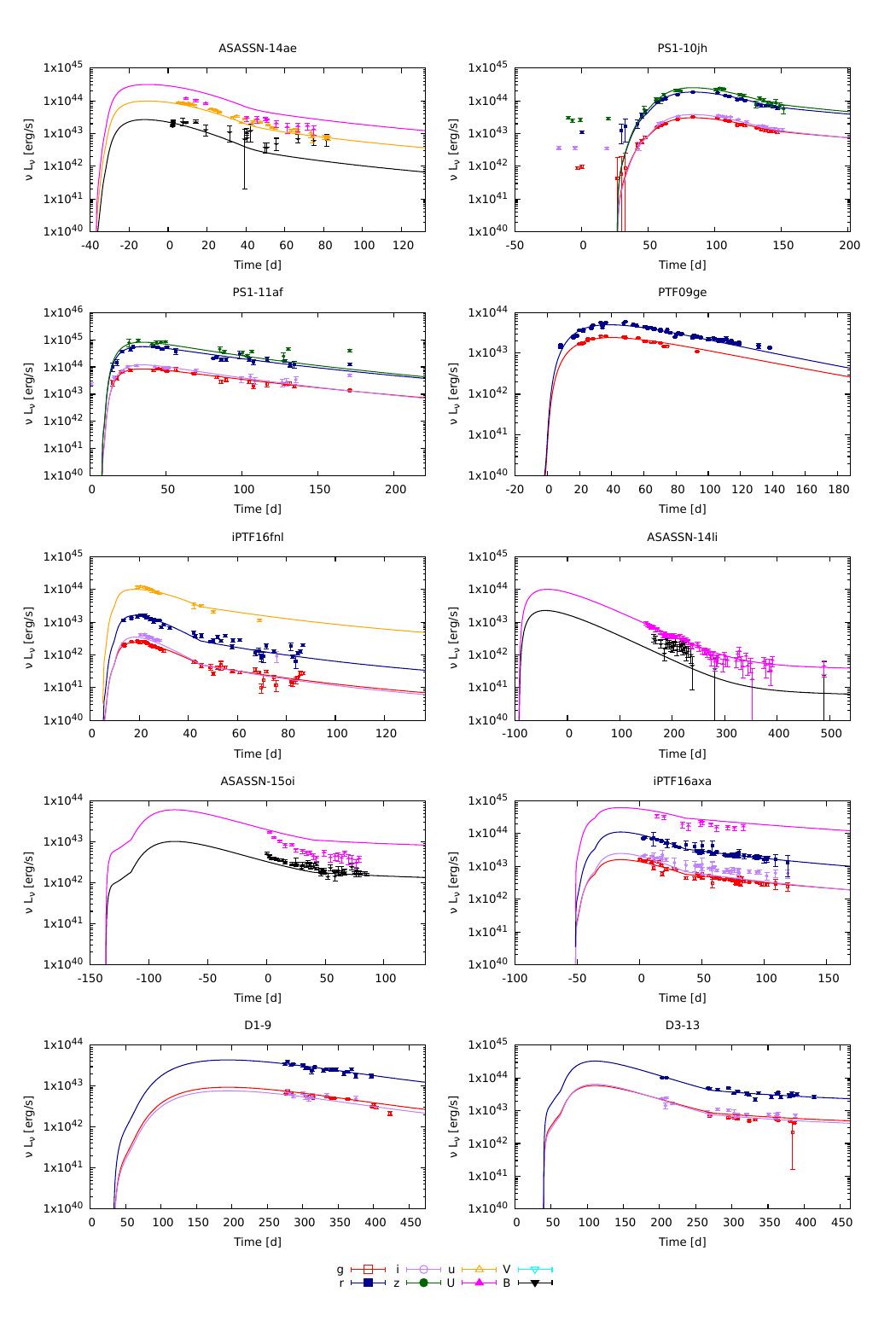}
    \caption{Light curves and their best-fit {\tt TiDE} models for the sample objects. The polytropic indices for the best-fit models were selected based on the p-value from the Kolmogorov-Smirnov test (see Tables~\ref{tab:4per3_res} and \ref{tab:5per3_res}). For clarity, only the model having the higher p-value is plotted in each panel. Different colors and symbols represent different bands (red empty square: $g$, blue filled square: $r$, purple empty circle: $i$, green filled circle: $z$, orange upward empty triangle: $u$, magenta upward filled triangle: $U$, cyan downward empty triangle: $V$, black downward filled triangle: $B$). The light curves in different bands were shifted vertically for better visibility by applying a constant multiplier of $+10^{0.5}$ with respect to the previous band.}
    \label{Fig:fitted_res1}
\end{figure*}

\begin{figure*}
    \centering
    \includegraphics[width=0.8\textwidth]{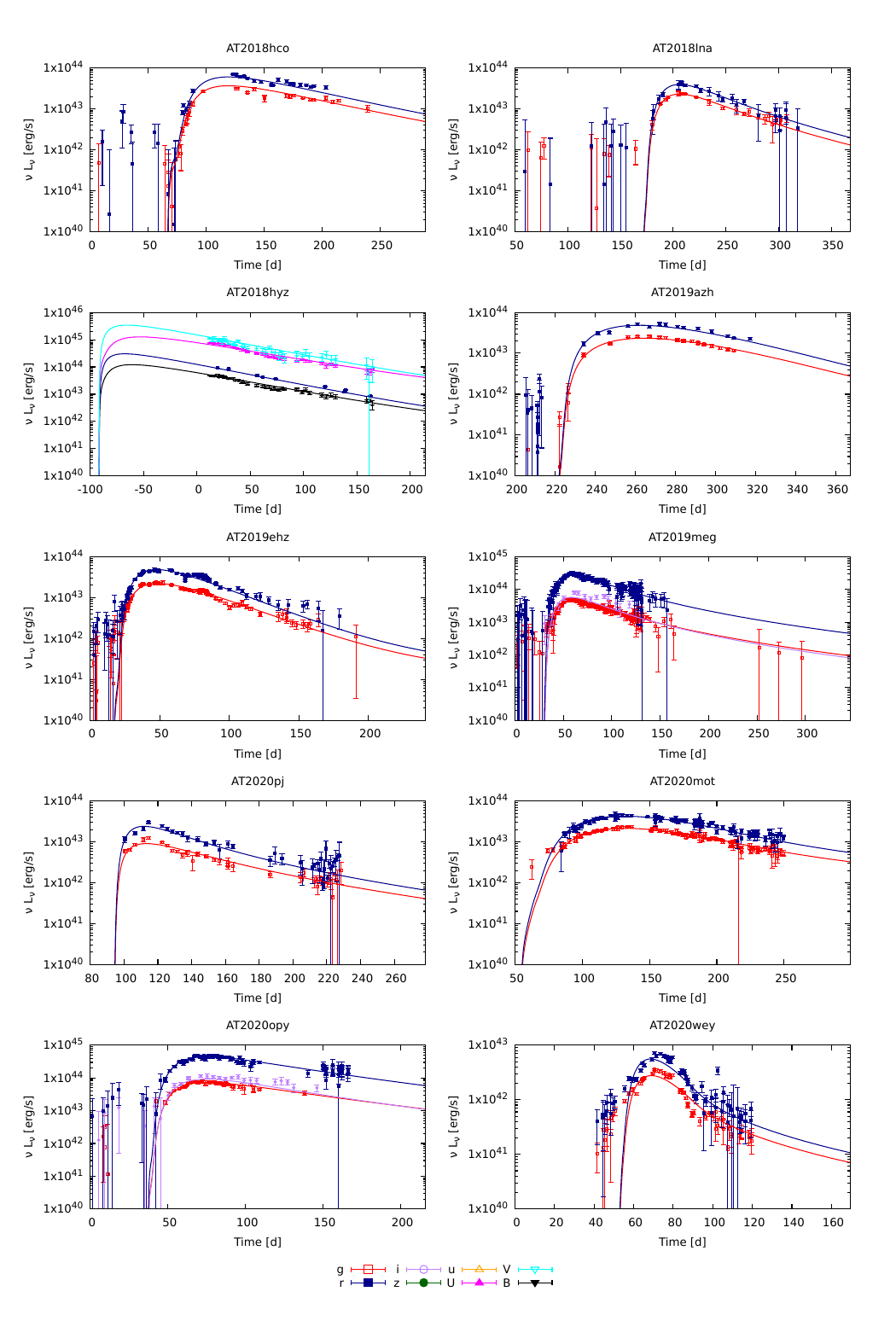}
    \caption{The same as Figure \ref{Fig:fitted_res1} but for the rest of the sample.}
    \label{Fig:fitted_res2}
\end{figure*}

Our best-fit models are compared to the observed light curves in Figure~\ref{Fig:fitted_res1} and \ref{Fig:fitted_res2}. In these plots different bands have different colors and symbols. For better comparison, the light curves from different bands were shifted vertically.

When applying {\tt TiDE}, the fitting parameters were the following ones: the initial time $t_{\rm ini}$ is the difference between our assumed $t=0$ and the moment of disruption (in days); $M_6$ is the mass of the SMBH (in $10^6 M_\odot$); $M_*$ is the mass of the star (in $M_\odot$), $\eta$ is the efficiency of converting the mass accretion rate to luminosity; $f_v$ is the ratio between the wind velocity and the escape velocity; $\epsilon_{\rm rep}$ is the efficiency of reprocessing the high-energy radiation from the disk to optical photons by the wind; and $t_{\rm diff}$ is the photon diffusion timescale in the optically thick wind (in days). The fitting range for each parameter was fixed except for $t_{\rm ini}$ that was adjusted to each particular object. The ranges are shown in Table \ref{tab:ranges}.
We fitted models having $\gamma = 4/3$ and $\gamma = 5/3$ polytropic indices separately, and the best-fit results for both models are collected in Table \ref{tab:4per3_res} and \ref{tab:5per3_res}.

\begin{table}
\centering
\caption{The fitting ranges}
\label{tab:ranges}
\begin{tabular}{ccc}
     \hline
     \hline
     Parameter & Min & Max\\
     \hline
     $M_6$ & $0.05$ & $100$\\
     $M_*$ & $0.05$ & $60$\\
     $\eta$ & $0.008$ & $0.4$\\
     $f_v$ & $1$ & $5$ \\
     $\epsilon_{\rm rep}$ & $0$ & $1$\\
     $t_{\rm diff}$ & $0$ & $100$\\
     \hline
\end{tabular}
\end{table}

We present the reduced $\chi^2$  values for each object alongside the fitted parameters in Tables~\ref{tab:4per3_res} and \ref{tab:5per3_res}. After having the models with the lowest $\chi^2$ for both polytropic indices, we chose the final best-fit model based on the $p$-value of the Kolmogorov-Smirnov test of the residuals \citep{2010arXiv1012.3754A}, and only the parameters of this final best-fit model are presented in Table~\ref{tab:obj_masses} and plotted in all figures.

\begin{figure*}
    \centering
    \includegraphics[width=1\textwidth]{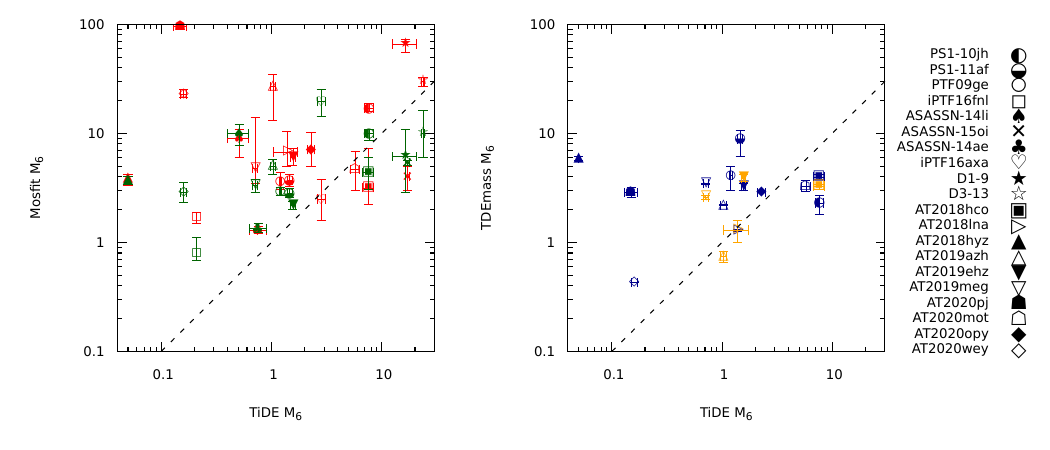}
    \caption{Comparison of the $M_6$ parameters from {\tt TiDE}, {\tt MOSFiT} and {\tt TDEmass}. Different colors correspond to the different codes (red: {\tt MOSFiT} 1, dark-green: {\tt MOSFiT} 2, dark-blue: {\tt TDEmass} 1, orange: {\tt TDEmass} 2). Indices represents the {\tt MOSFiT} and {\tt TDEmass} results from different papers (see Table \ref{tab:obj_masses}). The numerical values are collected in Table \ref{tab:obj_masses}}.
    \label{fig:M6_compare}
\end{figure*}

In Figure \ref{fig:M6_compare} we compare our new $M_6$ results from TiDE (plotted on the horizontal axis) to those from {\tt MOSFiT} (left) and {\tt TDEmass} (right) (shown on the vertical axis). The data are collected from Table \ref{tab:obj_masses}. Different events are plotted with different symbols, while different colors represent the different {\tt MOSFiT} and {\tt TDEmass} results based on different papers (red: first {\tt MOSFiT}, dark-green: second {\tt MOSFiT}, dark-blue: first {\tt TDEmass}, orange: second {\tt TDEmass}, for sources see Table \ref{tab:obj_masses}).  The dashed line illustrates the 1:1 relation.

It is seen that the $M_6$ values from the different methods do overlap each other in the $M_6\in[1;5]$ range, but outside that region {\tt TiDE} finds different values than the two other codes: for $M_6 < 1$ both {\tt MOSFiT} and {\tt TDEmass} predicts higher SMBH masses than {\tt TiDE}, while for $M_6 > 5$ it is the opposite. 
Note that this figure also illustrates the discrepancies between the {\tt MOSFiT} and {\tt TDEmass} results, and sometimes the two {\tt MOSFiT} results also differ from each other. Nevertheless it can be concluded that for most of the sample the {\tt TiDE} results are in an order-of-magnitude agreement with the previously published $M_6$ masses.

\begin{figure*}
    \centering
    \includegraphics[width=1\textwidth]{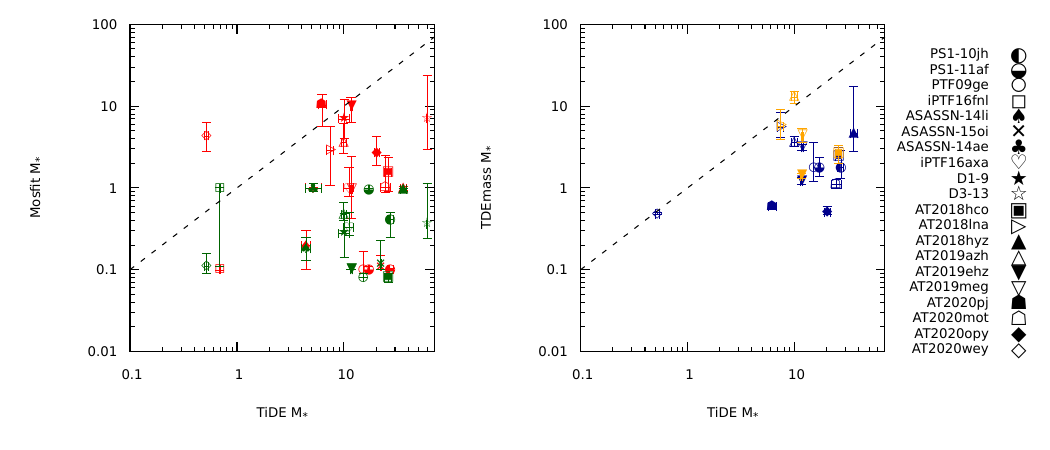}
    \caption{The same as Figure \ref{fig:M6_compare}, but for $M_*$.}
    \label{fig:Mstar_compare}
\end{figure*}

Figure \ref{fig:Mstar_compare} shows the same comparison, but for $M_*$ masses. It is seen that in most cases {\tt TiDE} found higher $M_*$ masses than the other two methods. It may also be interesting that for the majority of the sample {\tt MOSFiT} found sub-solar stellar masses, i.e. ($M_* \lesssim 1 M_\odot)$.


\section{Discussion}

The TDEs studied in this paper could be put into three groups based on the temporal coverage of their observed light curves. 

In the first category, the observation cover only the decaying part of the light curve: ASASSN-14ae, ASASSN-14li, ASASSN-15oi, iPTF16axa, D1-9, D3-13, AT2018hyz. In this case the moment of disruption has large uncertainty; it might be constrained only by previous non-detections (if those are available at all). All of the best-fit parameters may be in a wide range, and many combinations of them could be sufficiently describe the same light curve.

The second group of objects has an observed peak in their light curve. In this case the time of the event could be constrained into a well-defined range based on the relation $t_{\rm peak} = C \cdot t_{\rm min}$, where the $C$ factor depends on the polytropic index (see KV23). The allowed range for the $t_{\rm min}$ parameter could be determined based on the potential values of $M_6$ and $M_*$.

In the third group both the rising and the decaying parts of the light curve are well observed, and the duration of the rising part (the rise time) can be measured. In this case $t_{\rm min}$ can be calculated from the $t_{\rm min}$ - $t_{\rm peak}$ relation above, and the moment of the disruption can be tightly constrained. Note, however, that this simple picture is valid only for total disruptions, and details of the photon diffusion process may also affect the final results. 

In this study we fitted all objects assuming the disruption of a main sequence star, and do not attempt to model the observations with a white dwarf star. The reason for omitting compact stars from the models can be  justified by considering the time scales of the observed events. A white dwarf star is expected to have $\gamma=5/3$ polytropic index and a maximum mass of $\sim 1.4 M_\odot$. After modeling the possible $t_{\rm min}$ timescales for white dwarf disruptions with TiDE, we found that the maximum of the allowed
rise time
is $\sim 10$ days, which is much less than the rise time of our sample objects.

Based on our best fitting results (Table \ref{tab:obj_masses}) it can be seen that our black hole masses are mostly in between the previously published {\tt MOSFiT} and/or {\tt TDEmass} results. It is also conspicuous that our stellar masses are often high (in 7 out of 20 cases higher than $20 M_\odot$). Note, however, that in the contrary, {\tt MOSFiT} usually found low stellar masses (lower than $0.5 M_\odot$ in 18 out of 36 cases). 
Even though these extreme values might be due to selection effects, their distribution does not look realistic. The disagreement between the estimates from different codes suggests significant model dependency of the results.

It is also apparent that the best-fit $\epsilon_{\rm rep}$ parameters in {\tt TiDE} are usually low. To test the effect of this parameter on the model light curve, we attempt to fit some of the objects with $3$ different (fixed) $\epsilon_{\rm rep}$ parameters: $0.1$, $0.4$, $0.9$. We found that it resulted in parameters that are roughly consistent with each other. 
These findings strengthens the need for a reprocessing model that takes into account the non- or half-bound stellar debris as it was also suggested by \citet{2016MNRAS.455..859S} or \citet{Guillochon14}.

The differences between the best-fit results from the three modeling codes could be due to multiple reasons. {\tt MOSFiT} is able to model the light curves assuming either partial or total disruptions, while {\tt TDEmass} and {\tt TiDE} can only apply the latter case. The physical picture within the model is significantly different in the case of {\tt TDEmass}, while the theoretical background in both {\tt MOSFiT} and {\tt TiDE} is similar. Nevertheless, {\tt MOSFiT} makes a soft cut off at the Eddington luminosity when estimating the possible accretion rates (see Section 2.1.1 in M19). This restriction strongly influences both the mass of the SMBH and the mass of the star, as these two parameters have the largest effect on the peak of the light curve.

It is also instructive to compare the two {\tt MOSFiT} results (from two different papers) to each other. In many cases it is seen that the two best-fit {\tt MOSFiT} models does not match within their error bars (e.g  iPTF16axa, D1-9, D3-13, AT2019azh, AT2020wey). This discrepancy could also be due to several things. In the Appendix of \citet{2022MNRAS.515.5604N} the authors present some possible reasons behind the differences between their results and those of M19. One of them is the different priors applied on the radiative efficiency $\epsilon$ parameter. Secondly, they used a different method for optimization ({\tt DYNESTY} instead of {\tt EMCEE}). Also, they recalibrated the UVOT photometry, which may also have an effect on the final, best-fit solution. 

These differences could be partly responsible for the discrepancy between the remaining {\tt MOSFiT} results, too. However, we mention another potential cause here. {\tt MOSFiT} uses two different stellar models having $\gamma = 4/3$ or $5/3$ polytropic indices. During the fitting procedure the polytropic parameter is chosen directly based on the mass of the star. Stars having masses in the range of $1 \leq M_* \leq 15$ are automatically assigned to have $\gamma = 4/3$, while $M_* \leq 0.3$ or $M_* \geq 22$ stars get $\gamma = 5/3$ polytropic index. In the transition region the code uses hybrid fallback functions (see M19). In those cases when the best-fit models have stellar masses that could be represented with either polytropic index, the best-fit black hole masses are in huge discrepancy (see e.g iPTF16axa, D1-9, D3-13, AT2019azh, AT2019ehz, AT2020wey). Thus, it is possible that the ambiguity of the $\gamma$ parameter may also lead to a model-dependent SMBH mass in the {\tt MOSFiT} solutions.  

Thus, it is concluded that although the physics behind both {\tt MOSFiT} and {\tt TiDE} is similar, there are also important differences between the implementations in the two codes, which could strongly affect the final, best-fit model parameters. Since both codes use many assumptions and approximations, it would be difficult, almost impossible to judge which one could be more realistic. Nevertheless, at least it should be kept in mind that the existing TDE light curve solutions and their parameters are significantly model-dependent.

{\tt TiDE} has the potential of adding improvements to its modules easily because of its object-oriented environment, which may give a chance for easy development. It also has the ability of testing new physical conceptions and comparing many different models under the same circumstances. 

\section{Summary}
In this paper we fitted $20$ well-observed TDE objects with the new {\tt TiDE} code. During the selection procedure the most important criterion was the availability of previously published fitting results with either the {\tt MOSFiT} and/or the {\tt TDEmass} codes. All observations were corrected to the galactic extinction and redshift. In most cases more than one {\tt MOSFiT} results were available for the same event.

We found that all 20 events could be modeled by the total disruption of a $0.5 < M_* < 50$ M$_\odot$ star by a SMBH.
Based on the timescale of the increasing part it can be said, that these selected objects could not be a TDE with a white dwarf event. 

We found that the best-fit mass parameters, $M_6$ and $M_*$ given by the three codes span a wide range, usually more than an order of magnitude. 
It might be partly caused by the different physical pictures behind the three codes.
Although {\tt MOSFiT} and {\tt TiDE} share the assumption that the debris circularizates rapidly and a circular accretion disk forms, they also have differences (e.g. a soft cut-off at the Eddington luminosity in {\tt MOSFiT}), which may also be responsible for the significant differences between the mass parameters.

Note that sometimes two {\tt MOSFiT} solutions on the same event found significantly different results. The reason behind that is not fully clear, but it might be due to the differences in the treatment of the data. 

Based on the results above it can be concluded that the mass estimates from TDE light curves may suffer from a significant model dependence. It highlights the importance of the comparison of the predictions from different models, and, even more, the necessity of the improvement of the physical models of TDEs.

The {\tt TiDE} code is solely based on physical equations and assumptions, and calibrated to numerical hydrodynamic simulations, which is an improvement from simple scaling of the hydro simulations. The whole picture will be improved in the future by the involvement of better physical description of the reprocessing, more realistic stellar structures as well as the description of partial disruption scenarios, which can further enhance our understanding of tidal disruption events.

\begin{acknowledgements}
This research is part of the project ``Transient Astrophysical Objects" GINOP 2.3.2-15-2016-00033 of the National Research, Development and Innovation Office (NKFIH), Hungary, funded by the European Union.  
\end{acknowledgements}

\bibliography{ms}

\begin{thebibliography}{}
\expandafter\ifx\csname natexlab\endcsname\relax\def\natexlab#1{#1}\fi
\providecommand{\url}[1]{\href{#1}{#1}}
\providecommand{\dodoi}[1]{doi:~\href{http://doi.org/#1}{\nolinkurl{#1}}}
\providecommand{\doeprint}[1]{\href{http://ascl.net/#1}{\nolinkurl{http://ascl.net/#1}}}
\providecommand{\doarXiv}[1]{\href{https://arxiv.org/abs/#1}{\nolinkurl{https://arxiv.org/abs/#1}}}

\bibitem[{{Andrae} {et~al.}(2010){Andrae}, {Schulze-Hartung}, \&
  {Melchior}}]{2010arXiv1012.3754A}
{Andrae}, R., {Schulze-Hartung}, T., \& {Melchior}, P. 2010, arXiv e-prints,
  arXiv:1012.3754, \dodoi{10.48550/arXiv.1012.3754}

\bibitem[{{Bonnerot} \& {Lu}(2020)}]{2020MNRAS.495.1374B}
{Bonnerot}, C., \& {Lu}, W. 2020, \mnras, 495, 1374,
  \dodoi{10.1093/mnras/staa1246}

\bibitem[{{Charalampopoulos} {et~al.}(2023){Charalampopoulos}, {Pursiainen},
  {Leloudas}, {Arcavi}, {Newsome}, {Schulze}, {Burke}, \&
  {Nicholl}}]{2022arXiv220912913C}
{Charalampopoulos}, P., {Pursiainen}, M., {Leloudas}, G., {et~al.} 2023, \aap,
  673, A95, \dodoi{10.1051/0004-6361/202245065}

\bibitem[{{Chatzopoulos} {et~al.}(2013){Chatzopoulos}, {Wheeler}, {Vinko},
  {Horvath}, \& {Nagy}}]{Minim}
{Chatzopoulos}, E., {Wheeler}, J.~C., {Vinko}, J., {Horvath}, Z.~L., \& {Nagy},
  A. 2013, \apj, 773, 76, \dodoi{10.1088/0004-637X/773/1/76}

\bibitem[{{Chen} {et~al.}(2009){Chen}, {Madau}, {Sesana}, \&
  {Liu}}]{2009ApJ...697L.149C}
{Chen}, X., {Madau}, P., {Sesana}, A., \& {Liu}, F.~K. 2009, \apjl, 697, L149,
  \dodoi{10.1088/0004-637X/697/2/L149}

\bibitem[{{Chen} {et~al.}(2011){Chen}, {Sesana}, {Madau}, \&
  {Liu}}]{2011ApJ...729...13C}
{Chen}, X., {Sesana}, A., {Madau}, P., \& {Liu}, F.~K. 2011, \apj, 729, 13,
  \dodoi{10.1088/0004-637X/729/1/13}

\bibitem[{{Dotan} \& {Shaviv}(2011)}]{2011MNRAS.413.1623D}
{Dotan}, C., \& {Shaviv}, N.~J. 2011, \mnras, 413, 1623,
  \dodoi{10.1111/j.1365-2966.2011.18235.x}

\bibitem[{{Frank} \& {Rees}(1976)}]{1976MNRAS.176..633F}
{Frank}, J., \& {Rees}, M.~J. 1976, \mnras, 176, 633,
  \dodoi{10.1093/mnras/176.3.633}

\bibitem[{{Gezari}(2021)}]{2021ARA&A..59...21G}
{Gezari}, S. 2021, \araa, 59, \dodoi{10.1146/annurev-astro-111720-030029}

\bibitem[{{Golightly} {et~al.}(2019{\natexlab{a}}){Golightly}, {Coughlin}, \&
  {Nixon}}]{Golightly_2}
{Golightly}, E. C.~A., {Coughlin}, E.~R., \& {Nixon}, C.~J. 2019{\natexlab{a}},
  \apj, 872, 163, \dodoi{10.3847/1538-4357/aafd2f}

\bibitem[{{Golightly} {et~al.}(2019{\natexlab{b}}){Golightly}, {Nixon}, \&
  {Coughlin}}]{Golightly_1}
{Golightly}, E.~C.~A., {Nixon}, C.~J., \& {Coughlin}, E.~R. 2019{\natexlab{b}},
  \apjl, 882, L26, \dodoi{10.3847/2041-8213/ab380d}

\bibitem[{{Gomez} \& {Gezari}(2023)}]{2023arXiv230214070G}
{Gomez}, S., \& {Gezari}, S. 2023, arXiv e-prints, arXiv:2302.14070,
  \dodoi{10.48550/arXiv.2302.14070}

\bibitem[{{Guillochon} {et~al.}(2014){Guillochon}, {Manukian}, \&
  {Ramirez-Ruiz}}]{Guillochon14}
{Guillochon}, J., {Manukian}, H., \& {Ramirez-Ruiz}, E. 2014, \apj, 783, 23,
  \dodoi{10.1088/0004-637X/783/1/23}

\bibitem[{{Guillochon} \& {Ramirez-Ruiz}(2013)}]{2013ApJ...767...25G}
{Guillochon}, J., \& {Ramirez-Ruiz}, E. 2013, \apj, 767, 25,
  \dodoi{10.1088/0004-637X/767/1/25}

\bibitem[{{Hammerstein} {et~al.}(2023){Hammerstein}, {van Velzen}, {Gezari},
  {Cenko}, {Yao}, {Ward}, {Frederick}, {Villanueva}, {Somalwar}, {Graham},
  {Kulkarni}, {Stern}, {Andreoni}, {Bellm}, {Dekany}, {Dhawan}, {Drake},
  {Fremling}, {Gatkine}, {Groom}, {Ho}, {Kasliwal}, {Karambelkar}, {Kool},
  {Masci}, {Medford}, {Perley}, {Purdum}, {van Roestel}, {Sharma}, {Sollerman},
  {Taggart}, \& {Yan}}]{ZTF_30_TDE}
{Hammerstein}, E., {van Velzen}, S., {Gezari}, S., {et~al.} 2023, \apj, 942, 9,
  \dodoi{10.3847/1538-4357/aca283}

\bibitem[{{Hills}(1975)}]{1975Natur.254..295H}
{Hills}, J.~G. 1975, \nat, 254, 295, \dodoi{10.1038/254295a0}

\bibitem[{{Holoien} {et~al.}(2014){Holoien}, {Prieto}, {Bersier}, {Kochanek},
  {Stanek}, {Shappee}, {Grupe}, {Basu}, {Beacom}, {Brimacombe}, {Brown},
  {Davis}, {Jencson}, {Pojmanski}, \& {Szczygie{\l}}}]{2014MNRAS.445.3263H}
{Holoien}, T.~W.~S., {Prieto}, J.~L., {Bersier}, D., {et~al.} 2014, \mnras,
  445, 3263, \dodoi{10.1093/mnras/stu1922}

\bibitem[{{Holoien} {et~al.}(2016){Holoien}, {Kochanek}, {Prieto}, {Stanek},
  {Dong}, {Shappee}, {Grupe}, {Brown}, {Basu}, {Beacom}, {Bersier},
  {Brimacombe}, {Danilet}, {Falco}, {Guo}, {Jose}, {Herczeg}, {Long},
  {Pojmanski}, {Simonian}, {Szczygie{\l}}, {Thompson}, {Thorstensen}, {Wagner},
  \& {Wo{\'z}niak}}]{2016MNRAS.455.2918H}
{Holoien}, T.~W.~S., {Kochanek}, C.~S., {Prieto}, J.~L., {et~al.} 2016, \mnras,
  455, 2918, \dodoi{10.1093/mnras/stv2486}

\bibitem[{{Ivanov} {et~al.}(2005){Ivanov}, {Polnarev}, \&
  {Saha}}]{2005MNRAS.358.1361I}
{Ivanov}, P.~B., {Polnarev}, A.~G., \& {Saha}, P. 2005, \mnras, 358, 1361,
  \dodoi{10.1111/j.1365-2966.2005.08843.x}

\bibitem[{{Komossa} {et~al.}(2008){Komossa}, {Zhou}, {Wang}, {Ajello}, {Ge},
  {Greiner}, {Lu}, {Salvato}, {Saxton}, {Shan}, {Xu}, \&
  {Yuan}}]{2008ApJ...678L..13K}
{Komossa}, S., {Zhou}, H., {Wang}, T., {et~al.} 2008, \apjl, 678, L13,
  \dodoi{10.1086/588281}

\bibitem[{{Kov{\'a}cs-Stermeczky} \& {Vink{\'o}}(2023)}]{TiDE}
{Kov{\'a}cs-Stermeczky}, Z.~V., \& {Vink{\'o}}, J. 2023, \pasp, 135, 034102,
  \dodoi{10.1088/1538-3873/acb9bb}

\bibitem[{{Li} {et~al.}(2015){Li}, {Naoz}, {Kocsis}, \&
  {Loeb}}]{2015MNRAS.451.1341L}
{Li}, G., {Naoz}, S., {Kocsis}, B., \& {Loeb}, A. 2015, \mnras, 451, 1341,
  \dodoi{10.1093/mnras/stv1031}

\bibitem[{{Lodato} {et~al.}(2009){Lodato}, {King}, \& {Pringle}}]{L09}
{Lodato}, G., {King}, A.~R., \& {Pringle}, J.~E. 2009, \mnras, 392, 332,
  \dodoi{10.1111/j.1365-2966.2008.14049.x}

\bibitem[{{Lodato} \& {Rossi}(2011)}]{Lodato_Rossi11}
{Lodato}, G., \& {Rossi}, E.~M. 2011, \mnras, 410, 359,
  \dodoi{10.1111/j.1365-2966.2010.17448.x}

\bibitem[{{MacLeod} {et~al.}(2016){MacLeod}, {Guillochon}, {Ramirez-Ruiz},
  {Kasen}, \& {Rosswog}}]{2016ApJ...819....3M}
{MacLeod}, M., {Guillochon}, J., {Ramirez-Ruiz}, E., {Kasen}, D., \& {Rosswog},
  S. 2016, \apj, 819, 3, \dodoi{10.3847/0004-637X/819/1/3}

\bibitem[{{MacLeod} {et~al.}(2013){MacLeod}, {Ramirez-Ruiz}, {Grady}, \&
  {Guillochon}}]{2013ApJ...777..133M}
{MacLeod}, M., {Ramirez-Ruiz}, E., {Grady}, S., \& {Guillochon}, J. 2013, \apj,
  777, 133, \dodoi{10.1088/0004-637X/777/2/133}

\bibitem[{{Magorrian} \& {Tremaine}(1999)}]{1999MNRAS.309..447M}
{Magorrian}, J., \& {Tremaine}, S. 1999, \mnras, 309, 447,
  \dodoi{10.1046/j.1365-8711.1999.02853.x}

\bibitem[{{Metzger} \& {Stone}(2016)}]{2016MNRAS.461..948M}
{Metzger}, B.~D., \& {Stone}, N.~C. 2016, \mnras, 461, 948,
  \dodoi{10.1093/mnras/stw1394}

\bibitem[{{Mockler} {et~al.}(2019){Mockler}, {Guillochon}, \&
  {Ramirez-Ruiz}}]{Mosfit}
{Mockler}, B., {Guillochon}, J., \& {Ramirez-Ruiz}, E. 2019, \apj, 872, 151,
  \dodoi{10.3847/1538-4357/ab010f}

\bibitem[{{Nicholl} {et~al.}(2022){Nicholl}, {Lanning}, {Ramsden}, {Mockler},
  {Lawrence}, {Short}, \& {Ridley}}]{2022MNRAS.515.5604N}
{Nicholl}, M., {Lanning}, D., {Ramsden}, P., {et~al.} 2022, \mnras, 515, 5604,
  \dodoi{10.1093/mnras/stac2206}

\bibitem[{{Nixon} \& {Coughlin}(2022)}]{2022ApJ...927L..25N}
{Nixon}, C.~J., \& {Coughlin}, E.~R. 2022, \apjl, 927, L25,
  \dodoi{10.3847/2041-8213/ac5118}

\bibitem[{{Nixon} {et~al.}(2021){Nixon}, {Coughlin}, \& {Miles}}]{nixon21}
{Nixon}, C.~J., {Coughlin}, E.~R., \& {Miles}, P.~R. 2021, \apj, 922, 168,
  \dodoi{10.3847/1538-4357/ac1bb8}

\bibitem[{{Piran} {et~al.}(2015){Piran}, {Svirski}, {Krolik}, {Cheng}, \&
  {Shiokawa}}]{2015ApJ...806..164P}
{Piran}, T., {Svirski}, G., {Krolik}, J., {Cheng}, R.~M., \& {Shiokawa}, H.
  2015, \apj, 806, 164, \dodoi{10.1088/0004-637X/806/2/164}

\bibitem[{{Planck Collaboration} {et~al.}(2020){Planck Collaboration},
  {Aghanim}, {Akrami}, {Ashdown}, {Aumont}, {Baccigalupi}, {Ballardini},
  {Banday}, {Barreiro}, {Bartolo}, {Basak}, {Battye}, {Benabed}, {Bernard},
  {Bersanelli}, {Bielewicz}, {Bock}, {Bond}, {Borrill}, {Bouchet}, {Boulanger},
  {Bucher}, {Burigana}, {Butler}, {Calabrese}, {Cardoso}, {Carron},
  {Challinor}, {Chiang}, {Chluba}, {Colombo}, {Combet}, {Contreras}, {Crill},
  {Cuttaia}, {de Bernardis}, {de Zotti}, {Delabrouille}, {Delouis}, {Di
  Valentino}, {Diego}, {Dor{\'e}}, {Douspis}, {Ducout}, {Dupac}, {Dusini},
  {Efstathiou}, {Elsner}, {En{\ss}lin}, {Eriksen}, {Fantaye}, {Farhang},
  {Fergusson}, {Fernandez-Cobos}, {Finelli}, {Forastieri}, {Frailis},
  {Fraisse}, {Franceschi}, {Frolov}, {Galeotta}, {Galli}, {Ganga},
  {G{\'e}nova-Santos}, {Gerbino}, {Ghosh}, {Gonz{\'a}lez-Nuevo}, {G{\'o}rski},
  {Gratton}, {Gruppuso}, {Gudmundsson}, {Hamann}, {Handley}, {Hansen},
  {Herranz}, {Hildebrandt}, {Hivon}, {Huang}, {Jaffe}, {Jones}, {Karakci},
  {Keih{\"a}nen}, {Keskitalo}, {Kiiveri}, {Kim}, {Kisner}, {Knox},
  {Krachmalnicoff}, {Kunz}, {Kurki-Suonio}, {Lagache}, {Lamarre}, {Lasenby},
  {Lattanzi}, {Lawrence}, {Le Jeune}, {Lemos}, {Lesgourgues}, {Levrier},
  {Lewis}, {Liguori}, {Lilje}, {Lilley}, {Lindholm}, {L{\'o}pez-Caniego},
  {Lubin}, {Ma}, {Mac{\'\i}as-P{\'e}rez}, {Maggio}, {Maino}, {Mandolesi},
  {Mangilli}, {Marcos-Caballero}, {Maris}, {Martin}, {Martinelli},
  {Mart{\'\i}nez-Gonz{\'a}lez}, {Matarrese}, {Mauri}, {McEwen}, {Meinhold},
  {Melchiorri}, {Mennella}, {Migliaccio}, {Millea}, {Mitra},
  {Miville-Desch{\^e}nes}, {Molinari}, {Montier}, {Morgante}, {Moss}, {Natoli},
  {N{\o}rgaard-Nielsen}, {Pagano}, {Paoletti}, {Partridge}, {Patanchon},
  {Peiris}, {Perrotta}, {Pettorino}, {Piacentini}, {Polastri}, {Polenta},
  {Puget}, {Rachen}, {Reinecke}, {Remazeilles}, {Renzi}, {Rocha}, {Rosset},
  {Roudier}, {Rubi{\~n}o-Mart{\'\i}n}, {Ruiz-Granados}, {Salvati}, {Sandri},
  {Savelainen}, {Scott}, {Shellard}, {Sirignano}, {Sirri}, {Spencer},
  {Sunyaev}, {Suur-Uski}, {Tauber}, {Tavagnacco}, {Tenti}, {Toffolatti},
  {Tomasi}, {Trombetti}, {Valenziano}, {Valiviita}, {Van Tent}, {Vibert},
  {Vielva}, {Villa}, {Vittorio}, {Wandelt}, {Wehus}, {White}, {White},
  {Zacchei}, \& {Zonca}}]{Planck18}
{Planck Collaboration}, {Aghanim}, N., {Akrami}, Y., {et~al.} 2020, \aap, 641,
  A6, \dodoi{10.1051/0004-6361/201833910}

\bibitem[{{Rees}(1988)}]{1988Natur.333..523R}
{Rees}, M.~J. 1988, \nat, 333, 523, \dodoi{10.1038/333523a0}

\bibitem[{{Ryu} {et~al.}(2020){Ryu}, {Krolik}, \& {Piran}}]{TDEmass}
{Ryu}, T., {Krolik}, J., \& {Piran}, T. 2020, \apj, 904, 73,
  \dodoi{10.3847/1538-4357/abbf4d}

\bibitem[{{Shiokawa} {et~al.}(2015){Shiokawa}, {Krolik}, {Cheng}, {Piran}, \&
  {Noble}}]{2015ApJ...804...85S}
{Shiokawa}, H., {Krolik}, J.~H., {Cheng}, R.~M., {Piran}, T., \& {Noble}, S.~C.
  2015, \apj, 804, 85, \dodoi{10.1088/0004-637X/804/2/85}

\bibitem[{{Steinberg} \& {Stone}(2022)}]{2022arXiv220610641S}
{Steinberg}, E., \& {Stone}, N.~C. 2022, arXiv e-prints, arXiv:2206.10641.
\newblock \doarXiv{2206.10641}

\bibitem[{{Stone} \& {Metzger}(2016)}]{2016MNRAS.455..859S}
{Stone}, N.~C., \& {Metzger}, B.~D. 2016, \mnras, 455, 859,
  \dodoi{10.1093/mnras/stv2281}

\bibitem[{{Strubbe} \& {Quataert}(2009)}]{Strubbe09}
{Strubbe}, L.~E., \& {Quataert}, E. 2009, \mnras, 400, 2070,
  \dodoi{10.1111/j.1365-2966.2009.15599.x}

\bibitem[{{van Velzen} {et~al.}(2020){van Velzen}, {Holoien}, {Onori}, {Hung},
  \& {Arcavi}}]{2020SSRv..216..124V}
{van Velzen}, S., {Holoien}, T. W.~S., {Onori}, F., {Hung}, T., \& {Arcavi}, I.
  2020, \ssr, 216, 124, \dodoi{10.1007/s11214-020-00753-z}

\bibitem[{{van Velzen} {et~al.}(2019{\natexlab{a}}){van Velzen}, {Stone},
  {Metzger}, {Gezari}, {Brown}, \& {Fruchter}}]{2019ApJ...878...82V}
{van Velzen}, S., {Stone}, N.~C., {Metzger}, B.~D., {et~al.}
  2019{\natexlab{a}}, \apj, 878, 82, \dodoi{10.3847/1538-4357/ab1844}

\bibitem[{{van Velzen} {et~al.}(2011){van Velzen}, {Farrar}, {Gezari},
  {Morrell}, {Zaritsky}, {{\"O}stman}, {Smith}, {Gelfand}, \&
  {Drake}}]{2011ApJ...741...73V}
{van Velzen}, S., {Farrar}, G.~R., {Gezari}, S., {et~al.} 2011, \apj, 741, 73,
  \dodoi{10.1088/0004-637X/741/2/73}

\bibitem[{{van Velzen} {et~al.}(2019{\natexlab{b}}){van Velzen}, {Gezari},
  {Cenko}, {Kara}, {Miller-Jones}, {Hung}, {Bright}, {Roth}, {Blagorodnova},
  {Huppenkothen}, {Yan}, {Ofek}, {Sollerman}, {Frederick}, {Ward}, {Graham},
  {Fender}, {Kasliwal}, {Canella}, {Stein}, {Giomi}, {Brinnel}, {van Santen},
  {Nordin}, {Bellm}, {Dekany}, {Fremling}, {Golkhou}, {Kupfer}, {Kulkarni},
  {Laher}, {Mahabal}, {Masci}, {Miller}, {Neill}, {Riddle}, {Rigault},
  {Rusholme}, {Soumagnac}, \& {Tachibana}}]{2019ApJ...872..198V}
{van Velzen}, S., {Gezari}, S., {Cenko}, S.~B., {et~al.} 2019{\natexlab{b}},
  \apj, 872, 198, \dodoi{10.3847/1538-4357/aafe0c}

\bibitem[{{van Velzen} {et~al.}(2021){van Velzen}, {Gezari}, {Hammerstein},
  {Roth}, {Frederick}, {Ward}, {Hung}, {Cenko}, {Stein}, {Perley}, {Taggart},
  {Foley}, {Sollerman}, {Blagorodnova}, {Andreoni}, {Bellm}, {Brinnel}, {De},
  {Dekany}, {Feeney}, {Fremling}, {Giomi}, {Golkhou}, {Graham}, {Ho},
  {Kasliwal}, {Kilpatrick}, {Kulkarni}, {Kupfer}, {Laher}, {Mahabal}, {Masci},
  {Miller}, {Nordin}, {Riddle}, {Rusholme}, {van Santen}, {Sharma}, {Shupe}, \&
  {Soumagnac}}]{2021ApJ...908....4V}
{van Velzen}, S., {Gezari}, S., {Hammerstein}, E., {et~al.} 2021, \apj, 908, 4,
  \dodoi{10.3847/1538-4357/abc258}

\bibitem[{{Wang} {et~al.}(2012){Wang}, {Zhou}, {Komossa}, {Wang}, {Yuan}, \&
  {Yang}}]{2012ApJ...749..115W}
{Wang}, T.-G., {Zhou}, H.-Y., {Komossa}, S., {et~al.} 2012, \apj, 749, 115,
  \dodoi{10.1088/0004-637X/749/2/115}

\bibitem[{{Wang} {et~al.}(2011){Wang}, {Zhou}, {Wang}, {Lu}, \&
  {Xu}}]{2011ApJ...740...85W}
{Wang}, T.-G., {Zhou}, H.-Y., {Wang}, L.-F., {Lu}, H.-L., \& {Xu}, D. 2011,
  \apj, 740, 85, \dodoi{10.1088/0004-637X/740/2/85}

\bibitem[{{Zhang}(2023)}]{2023ApJ...948...68Z}
{Zhang}, X. 2023, \apj, 948, 68, \dodoi{10.3847/1538-4357/acc182}

\end{thebibliography}

\section*{Appendix}

\begin{table*}
\movetabledown=9cm
\begin{rotatetable*}
\caption{The fitting results with {\tt TiDE} MS stars with $\gamma=4/3$ polytropic index}
\label{tab:4per3_res}
\begin{tabular}{ccccccccccc}
\hline
\hline
Object name & $T_0$ & $t_{\rm ini}$ & $M_6$ & $M_*$ & $\eta$ & $f_v$ & $\epsilon_{\rm rep}$ & $t_{\rm diff}$ & $\chi^2$ & p \\
\hline
PS1-10jh & $55305.55$ & $ 15.00 \pm 0.65 $ & $ 10.23 \pm 0.62 $ & $ 24.04 \pm 2.03 $ & $ 4.04 \ 10^{-3} \pm 1.68 \ 10^{-4} $ & $ 1.85 \pm 0.19 $ & $ 1.20 \ 10^{-4} \pm 1.75 \ 10^{-4} $ & $ 14.02 \pm 0.93 $ & $ 12.65 $ & $ 7.43 \ 10^{-3} $\\
PS1-11af & $55546$ & $ 0.32 \pm 0.64 $ & $ 3.05 \pm 0.18 $ & $ 20.62 \pm 0.54 $ & $ 2.01 \ 10^{-2} \pm 1.04 \ 10^{-3} $ & $ 4.99 \pm 0.13 $ & $ 1.05 \ 10^{-4} \pm 2.03 \ 10^{-3} $ & $ 3.15 \pm 0.61 $ & $ 6.77 $ & $ 8.69 \ 10^{-3} $\\
PTF09ge & $54950$ & $ -5.51 \pm 0.20 $ & $ 1.06 \pm 0.03 $ & $ 13.62 \pm 0.09 $ & $ 1.10 \ 10^{-2} \pm 4.25 \ 10^{-4} $ & $ 5.00 \pm 0.03 $ & $ 4.09 \ 10^{-5} \pm 1.67 \ 10^{-4} $ & $ 25.23 \pm 0.37 $ & $ 30.65 $ & $ 7.73 \ 10^{-21} $\\
iPTF16fnl & $57612.5$ & $ 2.26 \pm 1.14 $ & $ 0.21 \pm 0.00 $ & $ 0.69 \pm 0.01 $ & $ 1.06 \ 10^{-3} \pm 2.46 \ 10^{-5} $ & $ 1.01 \pm 0.01 $ & $ 1.80 \ 10^{-2} \pm 1.71 \ 10^{-3} $ & $ 0.04 \pm 0.93 $ & $ 4.10 $ & $ 1.38 \ 10^{-4} $\\
ASASSN-14li & $56851.25$ & $ -97.26 \pm 19.59 $ & $ 0.51 \pm 0.11 $ & $ 4.41 \pm 0.44 $ & $ 8.22 \ 10^{-3} \pm 6.77 \ 10^{-4} $ & $ 2.09 \pm 0.47 $ & $ 2.04 \ 10^{-1} \pm 6.07 \ 10^{-2} $ & $ 47.75 \pm 3.13 $ & $ 1.47 $ & $ 1.11 \ 10^{-1} $\\
ASASSN-15oi & $57269.53$ & $ -149.60 \pm 1.00 $ & $ 17.08 \pm 0.64 $ & $ 22.00 \pm 0.18 $ & $ 1.20 \ 10^{-2} \pm 5.03 \ 10^{-4} $ & $ 5.00 \pm 0.01 $ & $ 2.00 \ 10^{-1} \pm 6.07 \ 10^{-3} $ & $ 0.23 \pm 0.01 $ & $ 15.58 $ & $ 4.71 \ 10^{-9} $\\
ASASSN-14ae & $56712.5$ & $ -41.04 \pm 1.21 $ & $ 0.76 \pm 0.13 $ & $ 5.23 \pm 0.86 $ & $ 1.01 \ 10^{-3} \pm 5.50 \ 10^{-6} $ & $ 2.54 \pm 0.43 $ & $ 1.03 \ 10^{-1} \pm 1.02 \ 10^{-2} $ & $ 5.36 \pm 0.13 $ & $ 20.98 $ & $ 2.99 \ 10^{-22} $\\
iPTF16axa   & $57537.40$ & $ -129.20 \pm 4.21 $ & $ 3.03 \pm 0.28 $ & $ 25.13 \pm 1.35 $ & $ 2.07 \ 10^{-3} \pm 6.08 \ 10^{-5} $ & $ 1.34 \pm 0.14 $ & $ 2.59 \ 10^{-4} \pm 2.49 \ 10^{-4} $ & $ 0.31 \pm 0.03 $ & $ 4.52 $ & $ 6.41 \ 10^{-6} $\\
D1-9        & $52954.36$ & $ -14.94 \pm 26.09 $ & $ 69.05 \pm 8.57 $ & $ 16.56 \pm 2.36 $ & $ 2.22 \ 10^{-1} \pm 6.19 \ 10^{-2} $ & $ 3.24 \pm 0.98 $ & $ 7.07 \ 10^{-1} \pm 2.37 \ 10^{-1} $ & $ 65.63 \pm 9.47 $ & $ 3.18 $ & $ 1.31 \ 10^{-3} $\\
D3-13       & $52815.57$ & $ -19.55 \pm 2.09 $ & $ 33.41 \pm 1.96 $ & $ 59.91 \pm 1.17 $ & $ 1.43 \ 10^{-2} \pm 8.16 \ 10^{-4} $ & $ 3.08 \pm 0.07 $ & $ 1.39 \ 10^{-3} \pm 1.79 \ 10^{-2} $ & $ 2.31 \pm 1.81 $ & $ 3.04 $ & $ 2.35 \ 10^{-3} $\\
AT2018hco   & $58300$ & $ 55.35 \pm 0.52 $ & $ 10.29 \pm 0.40 $ & $ 16.47 \pm 0.59 $ & $ 7.02 \ 10^{-2} \pm 7.73 \ 10^{-3} $ & $ 4.36 \pm 0.31 $ & $ 4.71 \ 10^{-5} \pm 1.22 \ 10^{-2} $ & $ 4.03 \pm 0.40 $ & $ 5.41 $ & $ 2.45 \ 10^{-3} $\\
AT2018lna   & $58300$ & $ 166.90 \pm 1.34 $ & $ 1.38 \pm 0.35 $ & $ 7.41 \pm 0.50 $ & $ 4.28 \ 10^{-2} \pm 1.61 \ 10^{-2} $ & $ 4.10 \pm 0.85 $ & $ 2.02 \ 10^{-1} \pm 6.96 \ 10^{-2} $ & $ 10.37 \pm 2.57 $ & $ 0.82 $ & $ 1.24 \ 10^{-1} $\\
AT2018hyz & $58420$ & $ -92.64 \pm 3.85 $ & $ 0.05 \pm 0.00 $ & $ 35.81 \pm 2.48 $ & $ 3.98 \ 10^{-1} \pm 1.41 \ 10^{-2} $ & $ 1.12 \pm 0.02 $ & $ 2.10 \ 10^{-1} \pm 1.94 \ 10^{-2} $ & $ 49.83 \pm 0.69 $ & $ 3.78 $ & $ 2.92 \ 10^{-5} $\\
AT2019azh   & $58300$ & $ 216.10 \pm 0.43 $ & $ 1.03 \pm 0.06 $ & $ 9.94 \pm 0.58 $ & $ 1.51 \ 10^{-3} \pm 3.36 \ 10^{-4} $ & $ 4.99 \pm 0.34 $ & $ 1.12 \ 10^{-1} \pm 1.61 \ 10^{-2} $ & $ 30.24 \pm 2.55 $ & $ 2.35 $ & $ 7.80 \ 10^{-4} $\\
AT2019ehz & $58570$ & $ 12.14 \pm 0.31 $ & $ 1.56 \pm 0.07 $ & $ 11.65 \pm 0.25 $ & $ 4.89 \ 10^{-3} \pm 1.64 \ 10^{-4} $ & $ 5.00 \pm 0.06 $ & $ 2.58 \ 10^{-4} \pm 3.17 \ 10^{-4} $ & $ 8.00 \pm 0.18 $ & $ 3.25 $ & $ 8.15 \ 10^{-10} $\\
AT2019meg   & $58550$ & $ 24.09 \pm 0.38 $ & $ 1.56 \pm 0.10 $ & $ 12.70 \pm 0.48 $ & $ 1.70 \ 10^{-2} \pm 1.22 \ 10^{-3} $ & $ 4.07 \pm 0.14 $ & $ 4.77 \ 10^{-5} \pm 4.00 \ 10^{-3} $ & $ 7.37 \pm 0.59 $ & $ 1.45 $ & $ 3.97 \ 10^{-1} $\\
AT2020pj & $58750$ & $ 90.93 \pm 1.53 $ & $ 0.12 \pm 0.01 $ & $ 4.40 \pm 0.19 $ & $ 6.67 \ 10^{-3} \pm 5.40 \ 10^{-4} $ & $ 3.68 \pm 0.25 $ & $ 3.02 \ 10^{-2} \pm 5.14 \ 10^{-3} $ & $ 9.98 \pm 1.44 $ & $ 1.65 $ & $ 2.58 \ 10^{-5} $\\
AT2020mot & $58940$ & $ 48.47 \pm 0.67 $ & $ 8.96 \pm 0.36 $ & $ 38.12 \pm 1.11 $ & $ 3.40 \ 10^{-3} \pm 1.16 \ 10^{-4} $ & $ 4.99 \pm 0.14 $ & $ 1.84 \ 10^{-5} \pm 3.56 \ 10^{-4} $ & $ 49.19 \pm 0.72 $ & $ 1.49 $ & $ 4.79 \ 10^{-1} $\\
AT2020opy & $58990$ & $ 30.84 \pm 0.47 $ & $ 3.36 \pm 0.19 $ & $ 24.75 \pm 1.11 $ & $ 1.52 \ 10^{-2} \pm 1.43 \ 10^{-3} $ & $ 4.99 \pm 0.21 $ & $ 2.61 \ 10^{-3} \pm 5.74 \ 10^{-3} $ & $ 10.92 \pm 0.57 $ & $ 1.66 $ & $ 2.75 \ 10^{-6} $\\
AT2020wey & $59080$ & $ 48.97 \pm 0.41 $ & $ 0.16 \pm 0.01 $ & $ 0.52 \pm 0.02 $ & $ 1.03 \ 10^{-3} \pm 1.58 \ 10^{-5} $ & $ 1.00 \pm 0.02 $ & $ 5.13 \ 10^{-2} \pm 4.82 \ 10^{-3} $ & $ 6.27 \pm 0.33 $ & $ 9.61 $ & $ 7.95 \ 10^{-11} $\\ \hline
\end{tabular}
\end{rotatetable*}
\end{table*}

\begin{table*}
\movetabledown=9cm
\begin{rotatetable*}
\caption{The fitting results with {\tt TiDE} MS stars with $\gamma=5/3$ polytropic index}
\label{tab:5per3_res}
\begin{tabular}{ccccccccccc}
\hline
\hline
Object name & $T_0$ & $t_{\rm ini}$ & $M_6$ & $M_*$ & $\eta$ & $f_v$ & $\epsilon_{\rm rep}$ & $t_{\rm diff}$ & $\chi^2$ & p \\
\hline
PS1-10jh & $55305.55$ & $ -2.33 \pm 0.83 $ & $ 7.66 \pm 0.66 $ & $ 27.12 \pm 1.73 $ & $ 2.85 \ 10^{-3} \pm 9.70 \ 10^{-5} $ & $ 1.99 \pm 0.07 $ & $ 1.33 \ 10^{-2} \pm 3.73 \ 10^{-3} $ & $ 8.10 \pm 1.66 $ & $ 12.69 $ & $ 2.05 \ 10^{-2} $\\
PS1-11af & $55546$ & $ -6.59 \pm 0.82 $ & $ 1.46 \pm 0.08 $ & $ 17.09 \pm 0.29 $ & $ 2.35 \ 10^{-2} \pm 2.06 \ 10^{-3} $ & $ 3.61 \pm 0.13 $ & $ 2.24 \ 10^{-3} \pm 1.33 \ 10^{-2} $ & $ 2.80 \pm 0.61 $ & $ 6.06 $ & $ 6.38 \ 10^{-2} $\\
PTF09ge & $54950$ & $ -16.07 \pm 0.33 $ & $ 1.20 \pm 0.03 $ & $ 15.09 \pm 0.04 $ & $ 8.49 \ 10^{-3} \pm 2.17 \ 10^{-4} $ & $ 5.00 \pm 0.02 $ & $ 2.91 \ 10^{-6} \pm 1.54 \ 10^{-4} $ & $ 16.77 \pm 0.25 $ & $ 28.33 $ & $ 4.06 \ 10^{-20} $\\
iPTF16fnl & $57612.5$ & $ 4.53 \pm 0.13 $ & $ 0.05 \pm 0.00 $ & $ 1.51 \pm 0.01 $ & $ 1.76 \ 10^{-2} \pm 1.02 \ 10^{-3} $ & $ 5.00 \pm 0.01 $ & $ 4.60 \ 10^{-5} \pm 1.48 \ 10^{-4} $ & $ 0.05 \pm 0.00 $ & $ 7.58 $ & $ 6.01 \ 10^{-6} $\\
ASASSN-14li & $56851.25$ & $ -58.65 \pm 35.49 $ & $ 0.63 \pm 0.67 $ & $ 3.25 \pm 0.83 $ & $ 8.20 \ 10^{-3} \pm 6.48 \ 10^{-3} $ & $ 2.89 \pm 0.90 $ & $ 2.09 \ 10^{-1} \pm 7.86 \ 10^{-2} $ & $ 48.50 \pm 9.22 $ & $ 1.52 $ & $ 1.04 \ 10^{-1} $\\
ASASSN-15oi & $57269.53$ & $ -150.00 \pm 0.52 $ & $ 9.65 \pm 0.31 $ & $ 18.94 \pm 0.18 $ & $ 6.89 \ 10^{-3} \pm 2.21 \ 10^{-4} $ & $ 5.00 \pm 0.01 $ & $ 2.00 \ 10^{-1} \pm 2.72 \ 10^{-3} $ & $ 0.19 \pm 0.00 $ & $ 16.95 $ & $ 5.18 \ 10^{-13} $\\
ASASSN-14ae & $56712.5$ & $ -43.78 \pm 1.33 $ & $ 0.25 \pm 0.01 $ & $ 1.74 \pm 0.07 $ & $ 1.01 \ 10^{-3} \pm 3.51 \ 10^{-6} $ & $ 1.00 \pm 0.03 $ & $ 2.95 \ 10^{-1} \pm 7.12 \ 10^{-3} $ & $ 4.86 \pm 0.16 $ & $ 21.30 $ & $ 1.97 \ 10^{-22} $\\
iPTF16axa   & $57537.40$ & $ -73.51 \pm 3.77 $ & $ 2.82 \pm 0.23 $ & $ 11.15 \pm 1.19 $ & $ 2.04 \ 10^{-3} \pm 4.08 \ 10^{-5} $ & $ 1.17 \pm 0.10 $ & $ 3.04 \ 10^{-5} \pm 1.77 \ 10^{-4} $ & $ 0.18 \pm 0.14 $ & $ 4.39 $ & $ 9.48 \ 10^{-6} $\\
D1-9        & $52954.36$ & $ -21.72 \pm 41.25 $ & $ 16.44 \pm 3.94 $ & $ 10.04 \pm 1.16 $ & $ 1.38 \ 10^{-1} \pm 4.81 \ 10^{-2} $ & $ 3.06 \pm 1.78 $ & $ 5.37 \ 10^{-1} \pm 4.64 \ 10^{-1} $ & $ 75.59 \pm 19.03 $ & $ 3.19 $ & $ 1.71 \ 10^{-3} $\\
D3-13       & $52815.57$ & $ -2.10 \pm 4.78 $ & $ 23.68 \pm 0.86 $ & $ 60.00 \pm 0.68 $ & $ 8.15 \ 10^{-3} \pm 3.87 \ 10^{-4} $ & $ 3.88 \pm 0.07 $ & $ 4.98 \ 10^{-4} \pm 1.42 \ 10^{-3} $ & $ 2.38 \pm 2.10 $ & $ 3.10 $ & $ 3.86 \ 10^{-3} $\\
AT2018hco   & $58300$ & $ 37.65 \pm 1.39 $ & $ 7.57 \pm 0.83 $ & $ 25.88 \pm 1.81 $ & $ 2.35 \ 10^{-2} \pm 2.97 \ 10^{-3} $ & $ 4.67 \pm 0.31 $ & $ 2.51 \ 10^{-4} \pm 3.54 \ 10^{-3} $ & $ 4.03 \pm 1.24 $ & $ 4.80 $ & $ 5.13 \ 10^{-3} $\\
AT2018lna   & $58300$ & $ 150.50 \pm 2.41 $ & $ 1.91 \pm 0.50 $ & $ 9.79 \pm 2.24 $ & $ 6.21 \ 10^{-3} \pm 4.45 \ 10^{-4} $ & $ 4.10 \pm 0.74 $ & $ 6.45 \ 10^{-2} \pm 1.73 \ 10^{-2} $ & $ 5.24 \pm 1.08 $ & $ 0.86 $ & $ 8.54 \ 10^{-2} $\\
AT2018hyz & $58420$ & $ -81.97 \pm 3.36 $ & $ 0.05 \pm 0.00 $ & $ 34.38 \pm 2.14 $ & $ 2.77 \ 10^{-1} \pm 1.13 \ 10^{-2} $ & $ 2.47 \pm 1.07 $ & $ 2.45 \ 10^{-4} \pm 3.15 \ 10^{-2} $ & $ 46.17 \pm 0.91 $ & $ 4.48 $ & $ 7.15 \ 10^{-7} $\\
AT2019azh   & $58300$ & $ 202.20 \pm 0.70 $ & $ 1.22 \pm 0.05 $ & $ 9.92 \pm 0.25 $ & $ 1.00 \ 10^{-3} \pm 4.32 \ 10^{-5} $ & $ 2.15 \pm 0.32 $ & $ 8.65 \ 10^{-2} \pm 6.79 \ 10^{-3} $ & $ 24.06 \pm 0.59 $ & $ 2.28 $ & $ 5.08 \ 10^{-5} $\\
AT2019ehz & $58570$ &$ -2.99 \pm 0.36 $ & $ 1.96 \pm 0.11 $ & $ 11.24 \pm 0.87 $ & $ 3.35 \ 10^{-3} \pm 8.84 \ 10^{-5} $ & $ 2.62 \pm 0.23 $ & $ 2.00 \ 10^{-5} \pm 4.01 \ 10^{-4} $ & $ 1.86 \pm 0.35 $ & $ 2.92 $ & $ 9.01 \ 10^{-11} $\\
AT2019meg   & $58550$ & $ 18.98 \pm 0.52 $ & $ 0.71 \pm 0.04 $ & $ 11.74 \pm 0.22 $ & $ 2.87 \ 10^{-2} \pm 2.52 \ 10^{-3} $ & $ 3.13 \pm 0.08 $ & $ 1.47 \ 10^{-4} \pm 8.05 \ 10^{-3} $ & $ 8.17 \pm 0.45 $ & $ 1.46 $ & $ 5.62 \ 10^{-1} $\\
AT2020pj & $58750$ &$ 88.04 \pm 0.75 $ & $ 0.15 \pm 0.02 $ & $ 6.24 \pm 0.53 $ & $ 7.46 \ 10^{-3} \pm 7.95 \ 10^{-4} $ & $ 4.67 \pm 0.41 $ & $ 3.73 \ 10^{-5} \pm 4.12 \ 10^{-3} $ & $ 5.70 \pm 0.73 $ & $ 1.37 $ & $ 8.15 \ 10^{-2} $\\
AT2020mot & $58940$ & $ 28.43 \pm 1.46 $ & $ 5.75 \pm 0.56 $ & $ 24.52 \pm 2.63 $ & $ 3.35 \ 10^{-3} \pm 1.04 \ 10^{-4} $ & $ 2.30 \pm 0.24 $ & $ 3.21 \ 10^{-4} \pm 3.17 \ 10^{-4} $ & $ 41.07 \pm 1.59 $ & $ 1.38 $ & $ 7.69 \ 10^{-1} $\\
AT2020opy & $58990$ & $ 20.23 \pm 0.93 $ & $ 2.28 \pm 0.18 $ & $ 20.25 \pm 1.19 $ & $ 1.72 \ 10^{-2} \pm 1.95 \ 10^{-3} $ & $ 3.45 \pm 0.21 $ & $ 9.72 \ 10^{-4} \pm 1.29 \ 10^{-2} $ & $ 7.62 \pm 0.88 $ & $ 1.57 $ & $ 7.44 \ 10^{-5} $\\
AT2020wey & $59080$ & $ 43.23 \pm 0.39 $ & $ 0.15 \pm 0.00 $ & $ 0.50 \pm 0.02 $ & $ 1.00 \ 10^{-3} \pm 5.08 \ 10^{-6} $ & $ 1.67 \pm 0.04 $ & $ 9.34 \ 10^{-1} \pm 3.01 \ 10^{-2} $ & $ 4.81 \pm 0.20 $ & $ 10.52 $ & $ 1.90 \ 10^{-12} $\\ \hline
\end{tabular}
\end{rotatetable*}
\end{table*}

\end{document}